\documentclass[12pt]{article}
\usepackage{graphicx,latexsym}

\def\a{\alpha} \def\b{\beta} \def\dl{\delta} \def\s{\sigma}  \def\vphi{\varphi} \def\eps{\epsilon}  \def\lam{\lambda} \def\Lam{\Lambda} \def\gm{\gamma} \def\Gm{\Gamma}   \def\nb{\nabla} \def\sq{\sqrt} 
 \def\pd{\partial} \def\pp{\prime} \def\half{\frac{1}{2}} \def\fr{\frac}

 \def\QG{{\rm QG}} \def\pl{{\rm pl}}         
   \def\V3{{\rm V}_3}

 \def\hg{{\hat g}}       \def\hG{{\hat G}}

\def\lap3{~| \!\!\! \partial^2} \def\dlap3{~| \!\!\! \partial^4}
\def\l:{: \!}\def\r:{\! :}

\def\lang{\langle} \def\rang{\rangle}
\def\barb{{\bar \beta}}

\begin{document}

\begin{titlepage}

\begin{flushright}
July 2014
\end{flushright}

\vspace{5mm}

\begin{center}
{\Large {\bf Renormalization Group Analysis for Quantum Gravity with A Single Dimensionless Coupling}}
\end{center}

\vspace{5mm}

\begin{center}
{\sc Ken-ji Hamada}\footnote{E-mail address: hamada@post.kek.jp; URL: http://research.kek.jp/people/hamada/}
\end{center}

\begin{center}
{\it Institute of Particle and Nuclear Studies, KEK, Tsukuba 305-0801, Japan} \\ and \\
{\it Department of Particle and Nuclear Physics, The Graduate University for Advanced Studies (Sokendai), Tsukuba 305-0801, Japan}
\end{center}

\begin{abstract}
We study the quantum conformal gravity whose dynamics is governed by a single dimensionless gravitational coupling with negative beta function. Since the Euler term is not dynamical classically, the constant in front of it is not an independent coupling. Quantum mechanically, however, it induces the Riegert conformal-factor dynamics with BRST conformal symmetry representing background free nature. In this paper, we propose how to handle the Euler term systematically, incorporating such dynamics on the basis of renormalization group analysis using dimensional regularization. As a nontrivial test of renormalization, we explicitly calculate the three-loop anomalous dimension of the cosmological constant operator and show that it agrees with the exact expression derived using the BRST conformal symmetry. The physical significance to inflation and the cosmic microwave background is also discussed.
\end{abstract}

\vspace{5mm}

\end{titlepage}

\section{Introduction}
\noindent

Recent observations of cosmic microwave background (CMB) anisotropies by various groups \cite{wmap,planck,bicep2} suggest that the Universe began to expand at very high energies beyond the Planck scale. There, spacetime would be totally fluctuating according to the laws of quantum mechanics. Thus, quantum gravity is now becoming an indispensable theory to describe the dynamics of the early Universe.

On the other hand, at first glance, it seems to be contradictory that we trace the origin of primordial fluctuations to quantum gravity, because the observations show that scalar fluctuations are more significant than tensor ones in the early Universe. This implies that if we wish to explain such fluctuations using quantum gravity only without adding a phenomenological scalar field, we have to construct a model whose physical states become scalarlike in the UV limit.

In order to resolve this problem, we propose a model of quantum gravity with a certain gauge symmetry imposing that all spacetimes related to each other under conformal transformations are equivalent, called the Becchi-Rouet-Stora-Tyutin (BRST) conformal invariance here, as a representation of the background-metric independence. It has been known that the Riegert action plays a significant role to realize it \cite{riegert,am,amm92,amm97,hs}. We then have shown that the Riegert theory including the kinetic term of the Weyl action has such a gauge symmetry as a part of diffeomorphism invariance, which is so strong that physical fields are restricted to ``real composite scalars" (called primary scalars) only \cite{hh, hamada12,hamada12RxS3}.\footnote{Due to the presence of this symmetry, the ghost modes in fourth-order gravitational fields, which are necessary for the conformal algebra to close, are not gauge invariant. So we will be able to begin an argument toward the resolution of the nonperturbative issue of unitarity.} 

The model we consider here is the renormalizable quantum theory of gravity expanded just from this background-free system by a single dimensionless coupling constant that brings the dynamics of traceless tensor fields.

The strategy we employ here to construct such a quantum gravity model is as follows. To begin with, we have to reduce the indefiniteness potentially existing in the fourth-order gravitational system. In order to settle this issue, we have considered massless QED in curved space as a gate for quantum gravity in view of the significance of conformal couplings. Since the QED-gravity coupling is unambiguously fixed by gauge symmetry and tractable much better than gravitational self-couplings, we can derive the definite conclusion on the gravitational counterterm at all orders \cite{hamada14}. Furthermore, we can expect the similarity between the gauge field part and the traceless tensor part ruled by the Weyl action. The quantum gravity model mentioned above can be constructed on the basis of this counterterm.

The original form of this quantum gravity model has been proposed in \cite{hamada02,hamada09}. The aim of this paper is to advance the study of this model based on the renormalization group (RG) equations and make sure that it is a consistent renormalizable quantum theory of gravity.

\section{Renormalization Group Analysis}
\noindent

From the analysis of the RG equations using dimensional regularization \cite{bc,hathrell-QED,hathrell-scalar,freeman,hamada14}, it has been recently shown that the gravitational counterterms for massless QED in curved space can be determined at all orders of the perturbation \cite{hamada14}, which are given by only two forms: the square of the Weyl tensor $C_{\mu\nu\lam\s}$ in $D$ dimensions and the modified Euler density defined by 
\begin{eqnarray}
   G_D = G_4 +(D-4) \chi(D) H^2 ,
      \label{expression of G_D}
\end{eqnarray} 
where $G_4$ is the usual Euler combination and $H=R/(D-1)$ is the rescaled scalar curvature. The coefficient $\chi$ is a finite function of $D$ only, which can be determined order by order in a series of $D-4$ as $\chi=\sum_{n=1}^\infty \chi_n (D-4)^{n-1}$ by solving the RG equations. The first three terms of $\chi$ are explicitly calculated as \cite{hamada14}
\begin{eqnarray}
     \chi_1=\half, \quad \chi_2 = \fr{3}{4}, \quad \chi_3 = \fr{1}{3} .
       \label{values of chi}
\end{eqnarray}

Furthermore, we have shown that the conformal anomaly related to the counterterm (\ref{expression of G_D}) is expressed in the form $E_D = G_D -4\chi \nb^2 H$. Here, it is significant that the familiar ambiguous term $\nb^2 R$ is fixed completely and this combination reduces to $E_4=G_4 -2 \nb^2 R/3$ in the four-dimensional limit due to $\chi_1=1/2$. It is just the combination proposed by Riegert, which produces the kinetic term of the conformal factor \cite{riegert}.

As for conformal couplings, we expect that the coefficients of $\chi$, especially $\chi_1$, have universal values independent of the theory. Really, the values (\ref{values of chi}) are not changed even when any number of fermions is added below, and also partially confirmed for $\lam \vphi^4$-scalar theory \cite{hathrell-scalar} and Yang-Mills theory \cite{freeman} in curved space.

On the other hand, unfortunately, we have found that the original action conjectured in the early work \cite{hamada02} as a renormalizable quantum theory of gravity becomes slightly different from the expression (\ref{expression of G_D}) at higher orders (see also footnote \ref{footnotee actually}). Therefore, it is necessary to reconsider our quantum gravity model based on the more proper action (\ref{expression of G_D}) determined in a theoretically established manner using the RG equations and check again whether it indeed gives a consistent quantum gravity.

The quantum gravity action we study here is thus defined by  
\begin{eqnarray*}
   S_g = \int d^D x \sq{g} \left\{ \fr{1}{t_0^2} C_{\mu\nu\lam\s}^2 + b_0 G_D \right\} 
\end{eqnarray*}
beyond the Planck scale, where $t_0$ is a dynamical coupling constant, while $b_0$ is not so, as discussed below. The lower-derivative actions such as the Einstein action are suppressed here.

We consider the perturbation theory in $t_0$ expanding about a conformally flat space defined by $C_{\mu\nu\lam\s}=0$, which is characterized by the expansion of the metric field:
\begin{eqnarray} 
    g_{\mu\nu} = e^{2\phi} {\bar g}_{\mu\nu}, \qquad 
    {\bar g}_{\mu\nu}=(\hg e^{t_0 h_0})_{\mu\nu} = \hg_{\mu\lam} ( \dl^\lam_{~\nu} + t_0 h^\lam_{0\nu} + \cdots ), 
      \label{metric expansion}
\end{eqnarray}
where $h^\lam_{0\lam}=0$ and $\hg_{\mu\nu}$ is the background metric. Thus, the quantum gravity model can be described as a quantum field theory on the background $\hg_{\mu\nu}$. At this time, it is significant that the conformal factor $e^{2\phi}$ is treated exactly without introducing its own coupling constant, because the conformally flat condition does not give any restrictions on it.

As in the previous study \cite{hamada14, hamada02}, we consider the model coupled to massless QED: $S=S_g + S_{\rm QED}$, where $S_{\rm QED} = \int d^D x \sq{g} \{ (1/4)F_{0\mu\nu}F_0^{\mu\nu} + \sum_{j=1}^{n_F} i{\bar \psi}_{0j} D\!\!\!\!/ \psi_{0j} \}$. The renormalization factors are defined by $A_{0\mu} = Z_3^{1/2} A_\mu$, $\psi_{0 j} = Z_2^{1/2} \psi_j$, and $h_{0\mu\nu}=Z_h^{1/2} h_{\mu\nu}$ for photon, $n_F$ fermions and traceless tensor fields, respectively, and $e_0 = \mu^{2-D/2} Z_3^{-1/2} e$ and $t_0= \mu^{2-D/2} Z_t t$ for coupling constants, where the Ward-Takahashi identity holds even when QED couples with quantized gravity. On the other hand, $\phi$ is not renormalized such that $Z_\phi=1$ because there is no coupling constant for this field.

The nonrenomalization theorem of $\phi$ is related to the geometrical property of $G_D$ (\ref{expression of G_D}) \cite{duff,ds}. Since its volume integral becomes topological at four dimensions, it is not dynamical at the classical level. Therefore, $b_0$ is not the independent coupling that governs the dynamics of gravity. So we expand the bare parameter $b_0$ in a pure-pole series as \cite{hamada02}
\begin{eqnarray*}
  b_0 = \fr{\mu^{D-4}}{(4\pi)^{D/2}} L_b, \qquad L_b = \sum_{n=1}^\infty \fr{b_n}{(D-4)^n} . 
\end{eqnarray*}
Since the field dependence of the volume integral of $G_D$ appears at $o(D-4)$, this expansion indicates that the dynamics is induced at the quantum level by canceling out the pole with the $D$ dependence of the action. The residues $b_n ~(n \geq 2)$ depend on the coupling constants, while the simple-pole residue has a coupling-independent part, which is divided as 
\begin{eqnarray*}
    b_1 = b + b_1^\pp , 
\end{eqnarray*}
where $b_1^\pp$ is coupling dependent and $b$ is a constant part.

In order to carry out the RG analysis systematically, incorporating the dynamics induced quantum mechanically, we propose the following procedure. For the moment, $b$ is regarded as a new coupling constant, instead of the finite term usually introduced in $b_0$. The effective action is then finite up to the topological term as follows:
\begin{eqnarray*}
    \Gm = \fr{\mu^{D-4}}{(4\pi)^{D/2}} \fr{b-b_c}{D-4} \int d^D x \sq{\hg} \hG_4 + \Gm_R(e,t,b) ,
\end{eqnarray*} 
where $\Gm_R$ is the renormalized quantity that depends on the coupling constants. The divergent term exists in a curved background only. The constant $b_c=11 n_F/360+40/9$ comes from the sum of direct one-loop calculations of QED \cite{duff} and gravitational fields \cite{ft,amm92,hs}.\footnote{In general, $b_c= (N_X + 11 N_F + 62 N_A)/360 + 769/180$, where $N_X$, $N_F$, and $N_A$ are the numbers of conformally coupled scalars, fermions, and gauge fields, and the last term is the sum of $-7/90$ and $87/20$ coming from the gravitational fields, $\phi$ and $h_{\mu\nu}$, respectively.} 
After the renormalization procedure is carried out, we take $b=b_c$. In this way, we can obtain the finite effective action $\Gm_R(e,t,b_c)$ whose dynamics is governed by a single gravitational coupling $t$.

The perturbative calculation can be performed by choosing the flat background as follows. Expanding the volume integral of $G_D$ (\ref{expression of G_D}), we obtain 
\begin{eqnarray*}
    && \int d^D x \sq{g} G_D = \sum_{n=0}^\infty \fr{(D-4)^n}{n!} \int d^D x \Bigl\{
                 4 \chi(D) \phi^n \pd^4 \phi  
           \nonumber \\
    &&  + (D-4) \zeta(D) \phi^n \pd^2 \phi \pd_\lam \phi \pd_\lam \phi
        + (D-4) \eta(D) \phi^n \left( \pd_\lam \phi \pd_\lam \phi \right)^2 
               \Bigr\} ,
\end{eqnarray*}
where $\zeta=8\chi-2(D-2)(D-3)$ and $\eta=(D-2)^2 \chi-(D-2)(D-3)^2$. The couplings with traceless tensor fields are not presented here (see \cite{hamada02}). As a convention in this paper, the same lower indices denote contraction by the flat background metric. Substituting the explicit values of $\chi_1$ and $\chi_2$ in (\ref{values of chi}) calculated using QED in curved space and neglecting $o((D-4)^3)$, the coefficients reduce to $\zeta=(8\chi_3-2)(D-4)^2$ and $\eta=(8\chi_3-1)(D-4)^2/2$. Therefore, the action is expanded as
\begin{eqnarray}
    && b_0 \int d^D x \sq{g} G_D = \fr{\mu^{D-4}}{(4\pi)^{D/2}} \int d^D x \biggl\{ 2b \phi \pd^4 \phi + (D-4) b \phi^2 \pd^4 \phi 
            \nonumber \\
    && + (D-4)^2 b \left[ \fr{1}{3} \phi^3 \pd^4 \phi + \half \left( 8 \chi_3 -1 \right) \left( \pd_\lam \phi \pd_\lam \phi \right)^2 \right] 
                + \cdots \biggr\} .
         \label{expansion of G_D action}
\end{eqnarray}
The first term of the rhs is just the Riegert action \cite{riegert} in flat background induced in terms of dimensional regularization, which gives the propagator of $\phi$. Thus, quantum corrections from this field are expanded in $1/b$, which corresponds to the large-$n_F$ expansion. The other terms are the induced vertices that become effective at loop level. The vertices from $\zeta$ are disregarded here because they are not the leading terms of each order of $D-4$ yielding the most singular diagrams. The terms denoted by dots include the vertices and counterterms coming from the residues $b_1^\pp$ and $b_n~(n \geq2)$. Similarly, the Weyl and QED actions are expanded in powers of $\phi$.

It has been checked in the original paper \cite{hamada02} that the renormalization procedure mentioned above, especially the nonrenormalization theorem of $\phi$, goes well at higher loops up to $o(e^6)$, $o(t^2)$, and $o(1/b)$ through direct calculations using the QED-gravity couplings and the gravitational self-couplings fixed by $\chi_{1,2}$ in (\ref{expansion of G_D action}) including the corresponding couplings with the traceless tensor field  not denoted here. In \cite{hamada02}, we have also discussed the RG equation for $b$, but it is somewhat incomplete. This paper aims to justify the procedure through the RG analysis and test it directly, including the vertices of higher orders coming from $\chi_3$. Therefore, we proceed with the argument without specifying the value of $\chi_3$ here.\footnote{\label{footnotee actually}Actually, the values of $\chi_{1,2}$ agree with those conjectured in \cite{hamada02}, but $\chi_3$ unfortunately disagrees.  In \cite{hamada02}, the $G_D$ action was determined by imposing the extra condition that the vertex $(\pd_\lam \phi \pd_\lam \phi)^2$ in (\ref{expansion of G_D action}) should vanish from analogy with the solvable two-dimensional quantum gravity action $\sq{g}R$ expanded near two dimensions. Although we imposed this condition with intent to rederive the exact expression of the anomalous dimension, we will find in Sect.4 that regardless of this vertex we can derive it.} 

The beta functions for the coupling constants $\a=e^2/4\pi$ and $\a_t=t^2/4\pi$ are defined by $\b = (\mu/\a) d \a/d\mu = D-4 + \barb$ and $\b_t = (\mu/\a_t) d \a_t/d\mu = D-4 + \barb_t$, respectively. Using the renormalization factors, they are expressed as $\barb= \mu d (\log Z_3)/d\mu$ and $\barb_t = \mu d(\log Z_t^{-2})/d\mu$.

The pure-pole term $L_b$ is divided as $L_b = b/(D-4) + L_b^\pp$. From the RG equation $\mu db_0/d\mu=0$, we obtain the expression
\begin{eqnarray*}
   \mu \fr{db}{d\mu} = (D-4) \barb_b , 
\end{eqnarray*}
where $\barb_b = - b -(D-4)L_b^\pp - \mu dL_b^\pp/d\mu$. In order that we can set $b$ to the constant $b_c$ at the end, this equation should satisfy the condition $\mu db/d\mu \to 0$ in the $D \to 4$ limit, and thus $\barb_b$ should be finite. Imposing this condition, we obtain the RG equation 
\begin{eqnarray*}
   \left( \a \fr{\pd}{\pd \a} + \a_t \fr{\pd}{\pd \a_t} + \barb_b \fr{\pd}{\pd b} + 1 \right) b_{n+1}
    + \left( \barb \a \fr{\pd}{\pd \a} + \barb_t \a_t \fr{\pd}{\pd \a_t} \right) b_n =0
\end{eqnarray*}
for $n \geq 1$, where $\pd b/\pd \a= \pd b/\pd \a_t =0$ is taken into account for $n=1$. This equation reduces to that in curved space \cite{hathrell-QED,hamada14} when the dependence on $\a_t$ and $b$ turns off. Thus, the present renormalization procedure is consistent with the results in curved space. The finite expression is then given by
\begin{eqnarray*}
    \barb_b = - \left( \fr{\pd b_1}{\pd b} \right)^{-1} 
        \left( b_1 + \a \fr{\pd b_1}{\pd \a} + \a_t \fr{\pd b_1}{\pd \a_t} \right).
\end{eqnarray*}
The $\a$-dependent terms of $b_1$ and $b_2$ have been calculated up to $o(\a^3)$ and $o(\a^4)$, respectively \cite{hathrell-QED, hamada14}. We only present $b_1 = b - (n_F^2/6)( \a/4\pi)^2$ and $b_2=(2 n_F^3/9)(\a/4\pi)^3$ here, and then obtain
\begin{eqnarray*}
   \barb_b = - b + \fr{n_F^2}{2} \left( \fr{\a}{4\pi} \right)^2 .
\end{eqnarray*}

The residues of $\log Z_3 = \sum_{n=1}^\infty f_n/(D-4)^n$ satisfy the RG equation 
\begin{eqnarray*}
    \a \fr{\pd f_{n+1}}{\pd \a} +\a_t \fr{\pd f_{n+1}}{\pd \a_t} + \barb_b \fr{\pd f_{n+1}}{\pd b} 
    + \barb \a \fr{\pd f_n}{\pd \a} + \barb_t \a_t \fr{\pd f_n}{\pd \a_t} = 0.
\end{eqnarray*}
From the direct loop calculations, the simple and double poles are given by \cite{hamada02} 
\begin{eqnarray*}
   f_1 &=& \fr{8 n_F}{3} \fr{\a}{4\pi} + \left( 4 n_F - \fr{16 n_F^2}{27b} \right) \left( \fr{\a}{4\pi} \right)^2,
              \nonumber \\
   f_2 &=& - \fr{32 n_F^2}{9} \left( \fr{\a}{4\pi} \right)^2 - \left( \fr{128 n_F^2}{9} - \fr{160 n_F^3}{81 b} \right)
             \left( \fr{\a}{4\pi} \right)^3 .
\end{eqnarray*}
Here, the corrections including $1/b$ come from the diagrams with an internal line of $\phi$. It has been shown that the $o(\a_t)$ correction to $f_1$ vanishes. Noting that $\barb_b=-b + o(\a^2)$, we find that $f_{1,2}$ are consistent with the RG equation. The beta function is now expressed as $\barb = \a \pd f_1/\pd \a + \a_t \pd f_1/\pd \a_t + \barb_b \pd f_1/\pd b$, and thus
\begin{eqnarray*}
    \barb = \fr{8 n_F}{3} \fr{\a}{4\pi} + \left( 8 n_F - \fr{16 n_F^2}{9}\fr{1}{b}  \right) \left( \fr{\a}{4\pi} \right)^2 .
\end{eqnarray*}
Replacing $b$ with $b_c$, we obtain the final expression of the beta function. The gravitational correction with $1/b_c$ gives a negative contribution.

In the same way, expanding as $\log Z_t^{-2} = \sum_{n=1}^\infty g_n/(D-4)^n$, we obtain the similar RG equations for $g_n$. The beta function has been calculated as \cite{ft,amm92,hs,hamada02}
\begin{eqnarray*}
    \barb_t = - \left( \fr{n_F}{20} + \fr{20}{3} \right) \fr{\a_t}{4\pi}
              - \fr{7 n_F}{36} \fr{\a \a_t}{(4\pi)^2} .
\end{eqnarray*}
Thus, the coupling constant $\a_t$ indicates the asymptotic freedom, which justifies the perturbation theory about conformally flat spacetime.

\section{Background-Metric Independence}
\noindent

The background-free nature of spacetime will be realized owing to the nonperturbative treatment of the conformal factor characterized by (\ref{metric expansion}). Indeed, the conformal change of the background metric $\hg_{\mu\nu} \to e^{2\s}\hg_{\mu\nu}$ is equivalent to the shift of the field $\phi \to \phi + \s$, which is nothing but the rewriting of the integration variable $\phi$. Therefore, the background-metric independence can be expressed as the shift invariance of the path integral measure. Thus, the background-metric independence is a purely quantum mechanical symmetry with respect to diffeomorphism realized when the metric field is quantized.

Reflecting the shift invariance, the background-metric independence is expressed as $\int D \phi \dl ( O e^{-S} )/\dl \phi =0$. Taking $\sq{g}\theta = \dl S/\dl \phi$ as an operator $O$, we obtain 
\begin{eqnarray}
   \lang \sq{g}\theta(x) \sq{g}\theta(y) \rang - \left\lang \fr{\dl \sq{g} \theta(x)}{\dl \phi(y)} \right\rang = 0 .
    \label{background free relation for theta}
\end{eqnarray}
Here, $\theta=\theta_A + \theta_\psi +\theta_g$ is merely the trace of the energy-momentum tensor. Each part is given by $\theta_A = (D-4) F_{0\mu\nu}F_0^{\mu\nu}/4$, $\theta_\psi = (D-1)\sum_{j=1}^{n_F} i {\bar \psi}_{0j} \!\! \stackrel{\leftrightarrow}{D\!\!\!\!/} \psi_{0j}$, and $\theta_g = (D-4) ( C_{\mu\nu\lam\s}^2/t_0^2 + b_0 E_D )$, where $E_D$ is the expression of the conformal anomaly mentioned before. The vanishing of the rhs is significant here, unlike the case in curved space discussed in \cite{hamada14}.

Let us directly see that Eq.(\ref{background free relation for theta}) indeed holds focusing the role of the $\phi$ field in the case of the flat background. To begin with, we expand $\sq{g}\theta$ in a series of $D-4$. The photon part is then expressed as 
\begin{eqnarray}
   \sq{g}\theta_A &=& (D-4) Z_3 \fr{1}{4} e^{(D-4)\phi} F_{\mu\nu}F_{\mu\nu} 
            \nonumber \\
    &=& \fr{D-4}{4} \left[ 1+ \fr{f_1}{D-4} + \left( D-4 + f_1 \right) \phi +\cdots \right] F_{\mu\nu}F_{\mu\nu} .
         \label{expansion of theta_A} 
\end{eqnarray}
The gravity part is expanded as
\begin{eqnarray}
   &&  \sq{g}\theta_g = \fr{\mu^{D-4}}{(4\pi)^{D/2}} \biggl\{ 4b \pd^4 \phi + (D-4)b \left[ 2\phi \pd^4 \phi + \pd^4 (\phi^2) \right]
          \nonumber \\
   &&  +(D-4)^2 b \left[ \phi^2 \pd^4 \phi + \fr{1}{3} \pd^4 (\phi^3) 
          -2 \left( 8\chi_3-1 \right) \pd_\lam \left( \pd_\lam \phi \pd_\s \phi \pd_\s \phi \right) \right] + \cdots \biggr\} ,
        \label{expansion of (D-4)b_0E_D}
\end{eqnarray}
where we write down the $b$-dependent leading terms only. The traceless tensor part in $\sq{g}\theta_g$ is expanded as $(D-4) e^{(D-4)\phi} {\bar C}^2_{\mu\nu\lam\s}/t_0^2$, where the Weyl tensor with the bar on it is defined in terms of the metric field ${\bar g}_{\mu\nu}$ in (\ref{metric expansion}). Since the traceless tensor part can be expanded as similar to (\ref{expansion of theta_A}), it is expected that this part can be explained as in the gauge field part shown below, although it is hard to handle self-interactions of the traceless tensor field directly.

\begin{figure}
\begin{center}
\includegraphics[width=10cm,angle=90]{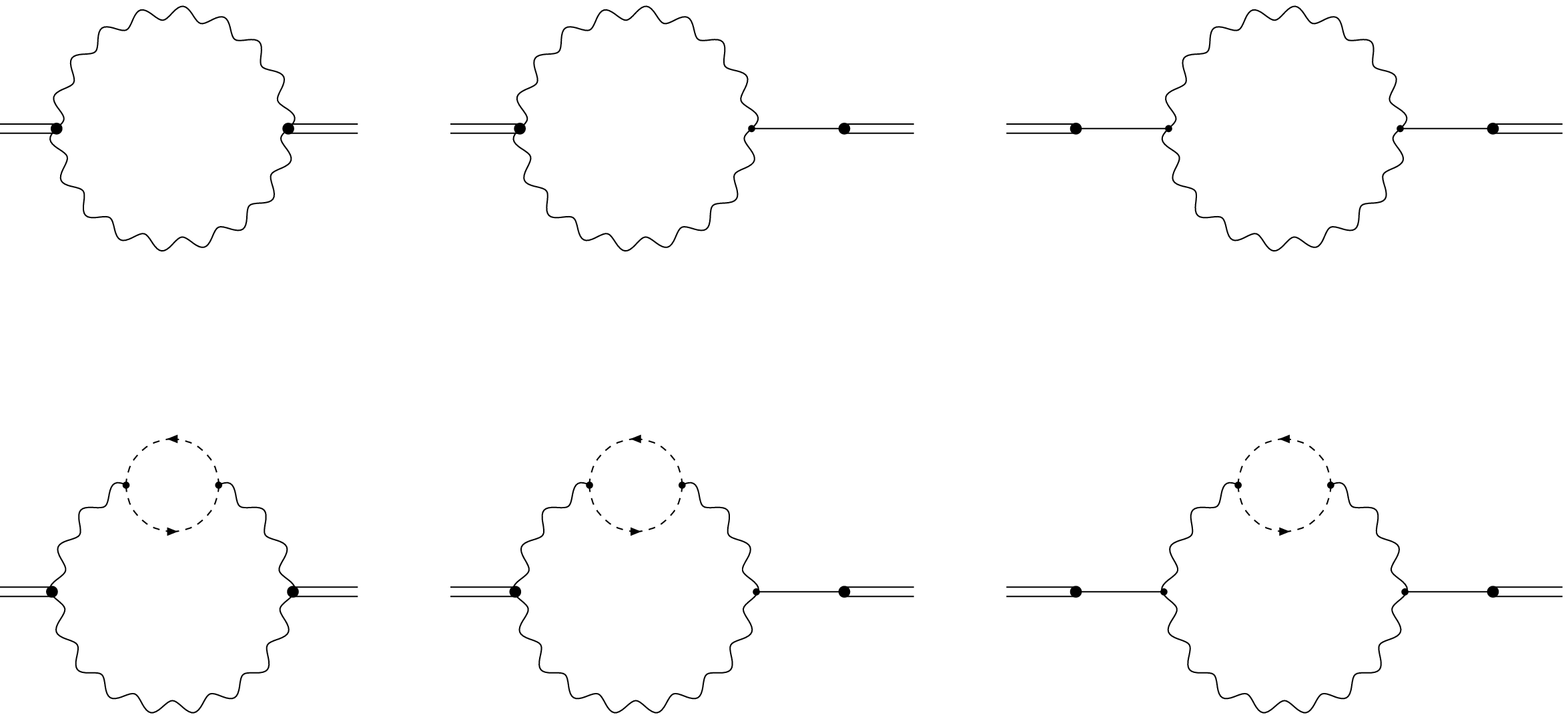}
\includegraphics[width=10cm,angle=90]{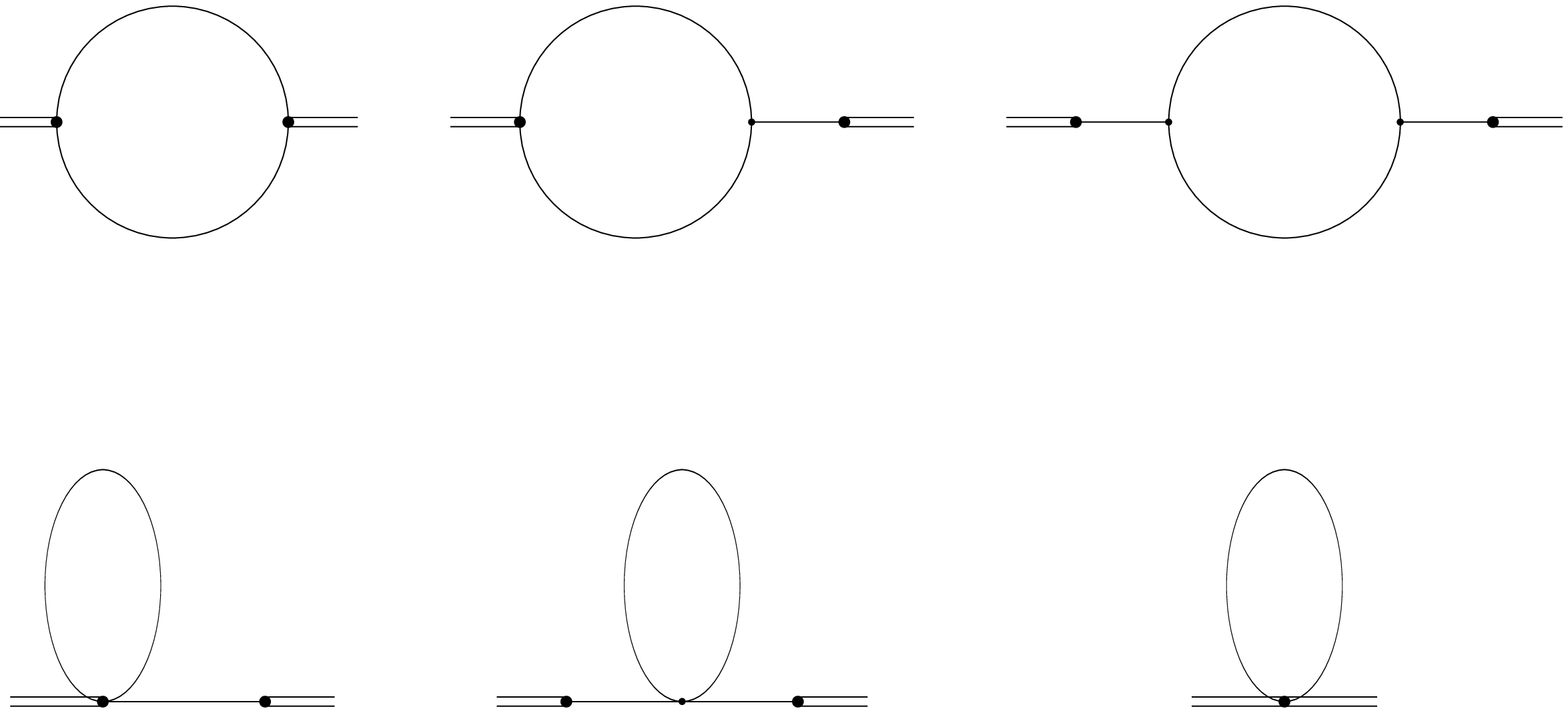}
\end{center}
\caption{\label{BMIfig} Diagrams representing the background-metric independence}
\end{figure}

The two-point function of $\sq{g}\theta$ at $o(b)$ can be easily calculated as $4b \mu^{D-4} p^4 /(4\pi)^{D/2}$ in momentum space from the first term of (\ref{expansion of (D-4)b_0E_D}) using the propagator $\lang \phi(p) \phi(-p) \rang = \mu^{4-D}(4\pi)^{D/2} /4b p^4$. It cancels out the second term of Eq.(\ref{background free relation for theta}).

Using (\ref{expansion of (D-4)b_0E_D}) and the vertices (\ref{expansion of G_D action}), we can show that Eq.(\ref{background free relation for theta}) is satisfied at $o(1)$ and $o(\a)$ as in Fig.\ref{BMIfig}, where the wavy, dashed-arrow, and solid lines denote the propagators of photon, fermions, and $\phi$, respectively, and the double line denotes $\sq{g}\theta$. The last three tadpole-type diagrams of $o(1)$ cannot be neglected in fourth-order theories unlike massless second-order theories (see Sect.4, for instance). The last one of these comes from the second term of Eq.(\ref{background free relation for theta}). At $o(\a)$, there is no contribution from the second term of Eq.(\ref{background free relation for theta}), which is consistent with the RG analysis deducing that there is no $o(\a)$-term in the residue $b_1$ \cite{hathrell-QED}. The result holds regardless of the value of $\chi_3$.

Finally, we see more explicitly that Eq.(\ref{background free relation for theta}) represents the background-metric independence. We here consider the energy-momentum tensor defined by the variation of the action with respect to the background metric as $\sq{\hg}{\hat \theta}= \dl S/\dl \s$, where $\dl/\dl\s = 2\hg_{\mu\nu}\dl/\dl \hg_{\mu\nu}$. Since the change of $\phi$ can be replaced with the change of $\s$ as far as the quantity is written in terms of the full metric (\ref{metric expansion}), the relationship $\sq{\hg}{\hat \theta}=\sq{g}\theta$ is satisfied, up to the gauge-fixing origin term that vanishes in physical correlation functions.  Therefore, Eq.(\ref{background free relation for theta}) can be expressed as $\dl^2 \lang 1 \rang /\dl \s(x) \dl \s(y)=0$, because the variation with respect to the background field $\s$ can be placed outside the correlation function. Thus, the model becomes independent of how to choose the background. It ensures that the calculation performed in the flat background is right.

\section{Cosmological Constant Operator}
\noindent

As a nontrivial test of the method proposed here, we calculate the anomalous dimension of the cosmological constant operator at $\a=\a_t=0$, for which the exact solution derived using the BRST conformal invariance has been known.

Since $\phi$ is not renormalized, the cosmological constant term $\Lam_0 \int d^D x \sq{g}$ can be simply renormalized by $\Lam_0 = \mu^{D-4} Z_\Lam \Lam$, where $\sq{g}=e^{D\phi}=\sum_{n=0}^\infty D^n \phi^n/n!$ in flat background. The anomalous dimension is defined by $\gm_\Lam =- (\mu/\Lam) d\Lam/d\mu$. Expanding the renormalization factor as $\log Z_\Lam = \sum_{n=1}^\infty u_n/(D-4)^n$, the anomalous dimension is given by $\gm_\Lam = D-4 +{\bar \gm}_\Lam$ and ${\bar \gm}_\Lam = \mu d(\log Z_\Lam)/d\mu = - b \pd u_1/\pd b$.

\begin{figure}
\begin{center}
\includegraphics[width=6cm]{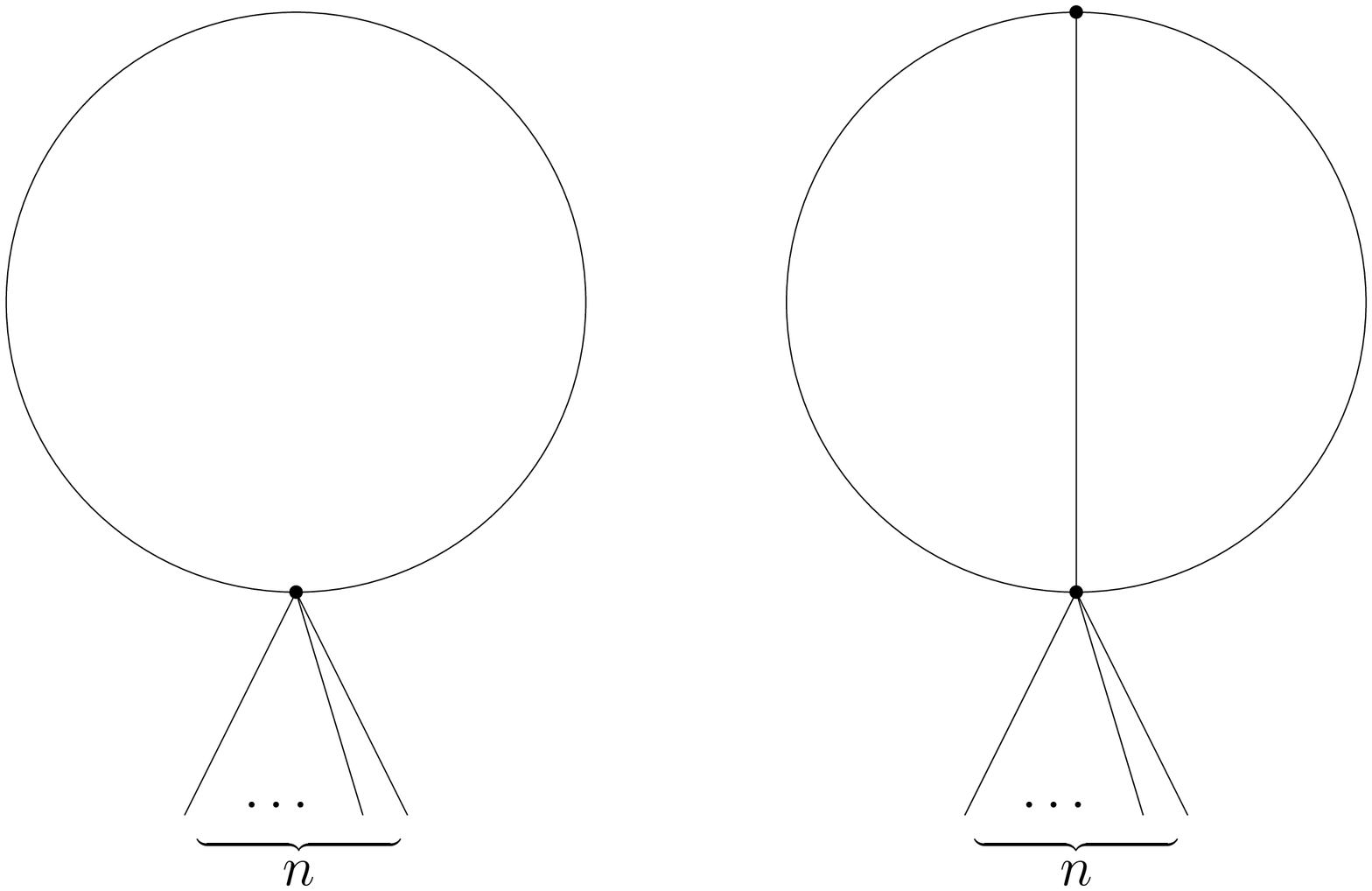}
\includegraphics[width=6cm]{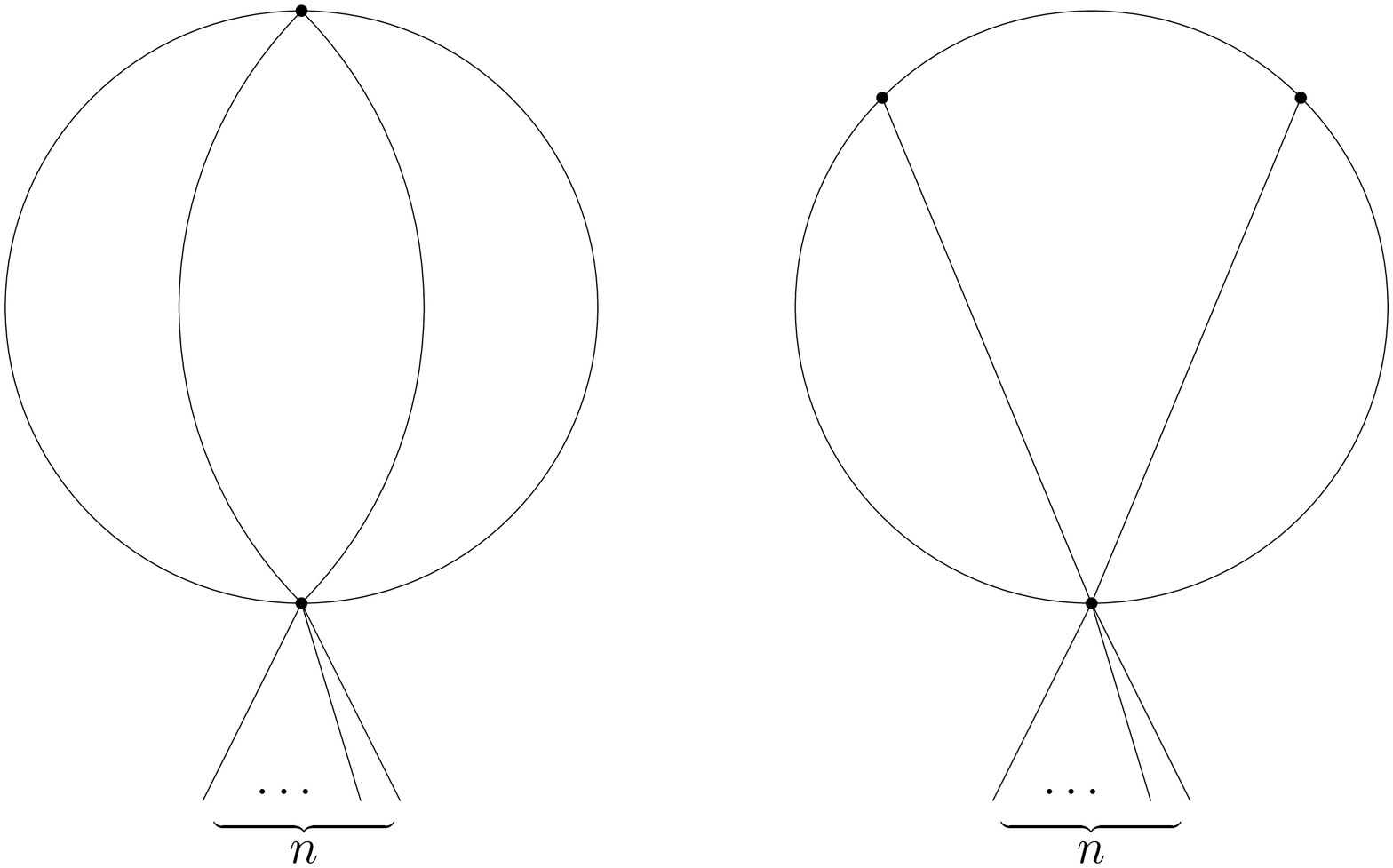}
\end{center}
\caption{\label{COSfig} Loop corrections to cosmological constant operator}
\end{figure}

The $o(1/b)$ and $o(1/b^2)$ corrections to $u_1$ are calculated from one- and two-loop diagrams in Fig.\ref{COSfig}, respectively. In order to handle IR divergences, we introduce an infinitesimal fictitious mass in the fourth-order propagator as $1/p^4 \to 1/p^4_z \equiv 1/(p^2 +z^2)^2$.\footnote{Since this mass is not gauge invariant, the $z$ dependences will be canceled out after summing up all contributions \cite{hamada09}. Incidentally, the lower-derivative action such as the Einstein action cannot be regarded as a usual mass term because there is an exponential conformal factor in this case.} 
Extracting UV divergences, we obtain $u_1=4/b+4/b^2$ \cite{hamada02}.

The $o(1/b^3)$ correction is obtained from two three-loop diagrams in Fig.\ref{COSfig} as
\begin{eqnarray}
    \left( -J + \fr{K}{8} \right) (4\pi)^{\fr{3D}{2}} \mu^{3(4-D)} \left( \fr{D}{4} \right)^4 \fr{(D-4)^2}{b^3} 
    \mu^{D-4} \Lam \fr{D^n}{n!} \int d^D x \phi^n ,
     \label{3-loop correction to cosmological constant}
\end{eqnarray} 
where $J$ and $K$, defined below, are the momentum integrals from the diagrams including a four-point vertex and two three-point vertices, respectively. The factor $(D-4)^2$ comes from these vertices in (\ref{expansion of G_D action}).

First, we calculate $J$, which is divided into two parts as $J=(1/3) J_A + (4\chi_3-1/2) J_B$ according to the two types of four-point vertex in (\ref{expansion of G_D action}). The integrals are then defined by 
\begin{eqnarray*}
    J_A &=& \fr{1}{4} \int dV_4 \fr{p^4+q^4+r^4+s^4}{p^4_z q^4_z r^4_z s^4_z} ,
         \nonumber \\
    J_B &=& \fr{1}{3} \int dV_4 
      \fr{(p \cdot q) (r \cdot s) + (p \cdot r) (q \cdot s) + (p \cdot s) (q \cdot r)}{p^4_z q^4_z r^4_z s^4_z}, 
\end{eqnarray*}
where $dV_4=[dpdqdrds] {\bar \dl}(p+q+r+s)$, $[dp]=d^D p/(2\pi)^D$ and ${\bar \dl}(p)=(2\pi)^D \dl^D (p)$. The most singular term of $J_A$ is simply given by the cube of the tadpole integral $I=\int [dp]/p^4_z =\Gm(2-D/2)z^{D-4}/(4\pi)^{D/2}$,\footnote{Since we focus on UV divergences, we can expand $p^4/p^4_z =1-2z^2/p^2+\cdots$ and disregard the $z$-dependent terms here.} 
and thus $J_A =I^3=1/(4\pi)^6 \eps^3$, where $D=4-2\eps$. The less singular terms and the IR divergence arising through $z^{3(D-4)}=1-6\eps \log z$ are neglected. Taking into account the overall factor $(D-4)^2$ in (\ref{3-loop correction to cosmological constant}), the contribution from $J_A$ gives a simple pole, which can be renormalized by taking $u_1=-8/3b^3$. On the other hand, the integral $J_B$ is given by
\begin{eqnarray*}
  J_B &=& \int [dk] \int [dp] \fr{p \cdot (p-k)}{p^4_z (p-k)^4_z} \int [dq] \fr{q \cdot (q+k)}{q^4_z (q+k)^4_z}
          \nonumber \\
      &=& \fr{z^{3(D-4)}}{(4\pi)^{\fr{3D}{2}}}\fr{1}{4} \Gm\left( 6-\fr{3D}{2} \right)
        \int^1_0 dx \int^1_0 dy \int^1_0 dt [x(1-x)y(1-y)]^{4-D} Y^{\fr{3D}{2}-6}
          \nonumber \\
      && \times
          [t(1-t)]^{2-\fr{D}{2}} \left\{ D^2 -2 D^2 (1-t) + D(D+2) t(1-t) \right\} ,
\end{eqnarray*}
where $Y=ty(1-y)+(1-t)x(1-x)$. This integral becomes finite apart from the factor $\Gm(6-3D/2)$, and so, multiplying $J_B$ by $(D-4)^2$, the simple pole vanishes. Thus, there is no contribution to $u_1$ from $J_B$, namely $\chi_3$.

The momentum integral $K$ is defined by
\begin{eqnarray*}
    K = \int dV_4 \fr{(p^4+r^4+(p+r)^4)(q^4+s^4+(p+r)^4)}{p^4_z q^4_z r^4_z s^4_z (p+r)^4_z} .
\end{eqnarray*}
As before, the integral is evaluated as $K=8 I^3 +{\tilde K}$, where $I$ is the tadpole integral and 
\begin{eqnarray*}
    {\tilde K} &=& \int [dk] k^4 \int \fr{[dp]}{p^4_z (p-k)^4_z} \int \fr{[dq]}{q^4_z (q+k)^4_z}
        \nonumber \\
    &=& \fr{z^{3(D-4)}}{(4\pi)^{\fr{3D}{2}}} \fr{D(D+2)}{4} \Gm\left( 6-\fr{3D}{2} \right)
        \int^1_0 dx \int^1_0 dy \int^1_0 dt 
           \nonumber \\
    &&   \times  
           [x(1-x)y(1-y)]^{3-D} [t(1-t)]^{3-\fr{D}{2}} Y^{\fr{3D}{2}-6} .
\end{eqnarray*}
Here, the multiple parameter integrals are numerically evaluated using the Monte Carlo method by Maple software and we find that ${\tilde K}$ has the triple-pole term as $(4\pi)^6 \eps^3 {\tilde K} =1.33$ at the $\eps \to 0$ limit.\footnote{Since the integrals of $x$ and $y$ are numerically dangerous, we change these variables as $dx/x(1-x)=dX$ such that $x(1-x)=1/4\cosh^2(X/2)$ and evaluate it within a finite range ($|X| \leq 8/\eps$ here), and similarly for $y$. Since the precision is not so good when $\eps$ is small, we evaluate $(4\pi)^{3D/2} \eps^3{\tilde K}$ (apart from overall $z$ dependence) at many values of $\eps$ more than $100$ between $0.02$ and $0.1$ and read off it as a polynomial function of $\eps$ using the least squares method. In this manner, we extract the value at $\eps \to 0$.} 
The numerical number is consistent with $4/3$. Therefore, substituting this value, we obtain $(4\pi)^6 \eps^3 K= 8+4/3 = 28/3$. Thus, the last diagram in Fig.\ref{COSfig} yields the simple pole $u_1=28/3b^3$.

Combining the results of $J$ and $K$, we obtain $u_1=20/3b^3$ as the three-loop correction. Thus, the anomalous dimension is given by
\begin{eqnarray*}
   {\bar \gm}_\Lam = \fr{4}{b} + \fr{8}{b^2} + \fr{20}{b^3} 
\end{eqnarray*}
with taking $b=b_c$ at last. This result agrees with the first three terms in the large-$b_c$ expansion of the exact expression ${\bar \gm}_\Lam=\gm-4=\gm^2/4b_c$, where $\gm=2b_c(1-\sq{1-4/b_c})$ is the charge for the cosmological constant operator defined by $\int d^4 x \l: e^{\gm\phi} \r:$ \cite{am, hh, hamada12,hamada12RxS3}.

In this way, we have seen that the renormalization manner proposed here can reproduce the exact anomalous dimension. Consequently, the result is independent of the value of $\chi_3$, while the values of $\chi_1$ and $\chi_2$ in (\ref{values of chi}) are crucial here. It suggests that vertices of the type $b(D-4)^n \phi^{n+1} \pd^4 \phi$ only contribute to $u_1$. We expect that $\chi_n$s are independent of the theory as mentioned before, but those of $n \geq 3$ may depend on it.

\section{Conclusion and Cosmological Implications}
\noindent

The goal of quantum gravity is to break the wall of Planck scale and to reveal the laws of physics there. We imagine that once we go beyond the Planck scale, there spreads a harmonious space without scale and singularity. The quantum gravity model with a single dimensionless gravitational coupling has a lot of desirable properties to describe such a spacetime. In this paper, we have studied the model by developing how to treat the conformal factor on the basis of the RG method using dimensional regularization, and then have seen that it passes several theoretical tests of renormalization.

The asymptotically free behavior of traceless tensor fields indicates that quantum fluctuations of the conformal factor become more significant than tensor fluctuations in the early epoch of the Universe. There, a novel spacetime phase with background-free nature called the BRST conformal symmetry will emerge. Such a phase can be imaged by a four-dimensional simplicial manifold with varying connectivity of simplices, called dynamical triangulation \cite{hey}. It provides a scale-invariant scalar spectrum primordially.

There are three gravitational mass scales in the model, namely the Planck mass $m_\pl$, the dynamical IR scale $\Lam_\QG$ indicated from the asymptotic freedom, and the cosmological constant. We set the ordering of the first two as $m_\pl \gg \Lam_\QG$ and the cosmological constant is considered to be negligible. In this case, the Riegert conformal dynamics is still significant about the Planck scale. The model then has a stable inflationary solution that can explain the results of CMB experiments well, whose evolution scenario is given as follows \cite{hy,hhy06, hhy10}. When the energy is going down to about $m_\pl$, the Einstein action comes into play and the Universe evolves to the inflationary phase. The inflation will terminate eventually at the dynamical scale $\Lam_\QG$ because a conformally flat spacetime is no longer significant there, and the Universe turns to the classical Einstein phase.\footnote{The low energy effective theory of gravity valid below the energy scale $\Lam_\QG$ will be given by an expansion in derivatives of the metric field, in which the Einstein action is dominated \cite{hhy06}.} 
The number of $e$-foldings is then given by the ratio of two mass scales as ${\cal N}_e \simeq m_\pl/\Lam_\QG$ and the amplitude of scalar fluctuation is roughly estimated to be $\dl R/R \simeq (\Lam_\QG/m_\pl)^2$. Thus, $\Lam_\QG$ is predicted to be the order of $10^{17}$GeV \cite{hy}. This scale can also explain the sharp falloff of CMB angular power spectra at very large angles \cite{hy, hhy10}.

In order to achieve the overall fit to the CMB data, more detailed consideration on the evolution process may be necessary. For a substance to make up the Universe, if there is a stable gravitational soliton, it will be a candidate for dark matter. They are left for future study.


\end{document}